\documentclass[11pt]{article}
\usepackage{latexsym}

\newcommand{\p}{\phi_}

\newcommand{\bq}{\begin{equation}} 
\newcommand{\eq}{\end{equation}} 
\newcommand{\bqali}{\begin{eqnarray}}
\newcommand{\eqali}{\end{eqnarray}} 
\newcommand{\no}{\noindent}
\newcommand{\g}[1]{\mathsf{#1}}

\title{Supersymmetric extensions of affine Toda theories}

\author{A. Opfermann\footnote{e-mail:ao200@damtp.cam.ac.uk}\\
\normalsize
\em Department of Applied Mathematics and Theoretical Physics, \\
\normalsize
\em University of Cambridge, \\
\normalsize
\em Silver Street, \\
\normalsize
\em Cambridge, CB3 9EW, UK}

\begin{document}
%%%%%%%%%%%%%%%%%%%%%%%%%%%%%%%%%%%%%%%%%%%%%%%%%%%%%%%%%%%%%%%%%%%%%%%%%%%%

\maketitle

\begin{abstract}

 It is shown  that all affine Toda
theories admit (1,0) supersymmetric extensions. The construction is based on
classical Lie algebras and  supersymmetric 
massive sigma models. The
supersymmetrized affine Toda theories have a unique,
supersymmetric vacuum, their mass matrix is well defined and their
energy functional is   positive semi-definite.

\bigskip

\noindent DAMTP-1998-95

\end{abstract}

\vfill

%%%%%%%%%%%%%%%%%%%%%%%%%%%%%%%%%%%%%%%%%%%%%%%%%%%%%%%%%%%%%%%%%%%%%%%%%

\section{Introduction}

Bosonic Toda theories \cite{botoda1, botoda2} 
provide a uniform way of constructing both conformally invariant and 
massive  field theories, which are integrable. 
Since the theory of Lie algebras underlies the integrability of the bosonic
models, it is natural to employ the theory of superalgebras 
to incorporate fermions into the bosonic models
\cite{sual1, sual2}.  Whereas with this method 
the classical integrability of the bosonic Toda theories is maintained,  
most of the fermionic extended
Toda theories  are plagued by an indefinite signature in their kinetic 
term \cite{holly} or by the
explicit breaking of their supersymmetry \cite{sual1}. 
Some notable exceptions, which
provide examples of  supersymmetric, unitary and  integrable field
theories, are the
supersymmetric extensions of Liouville theory \cite{lio1, lio2}, 
of sinh- and sine-Gordon
theory \cite{sine1, sine2} and of coupled  sinh- and sine-Gordon
theories \cite{coup}.

Toda theories are two-dimensional field theories with a flat target
space and an exponential  potential. Thus they constitute
examples of bosonic massive sigma models. With this in mind, 
it was recently
proposed  \cite{toda} to apply the framework of supersymmetric 
massive sigma models
\cite{mass1} to supersymmetrize
bosonic  Toda theories. In particular, 
(1,0) supersymmetric extensions were found 
for all conformal Toda theories
and for affine Toda theories based on the  algebra 
$A_r^{(1)}$. Attempts to construct further
supersymmetric extensions of these theories 
proved  unsuccessful, except for affine Toda theories based on an algebra of
rank one,
for which a (1,1) supersymmetric extension was given. 
On the one hand this reconfirms the
belief that there is no N=1 supersymmetric\footnote{In chiral
notation N=1 supersymmetry
corresponds to  (1,1) supersymmetry.} theory whose bosonic part is a
Toda model based on a simple Lie algebras of rank bigger than one
\cite{holly}. On the other hand it points to
previously overlooked (1,0) supersymmetric
Toda theories, which have the same a 
bosonic part as  a  bosonic Toda theory based on a
simple Lie algebra.

It is an interesting question, whether or not these (1,0)
supersymmetric extensions of Toda theories share the integrability
property with their bosonic counterparts. 
Underlying the classical integrability of the bosonic models is a Lax pair
or zero curvature conditions \cite{lax1, lax2}, which 
imply the existence of an infinite number of
conserved currents. Furthermore, it appears that their 
S-matrices   are exact \cite{sma1, sma3}, which renders 
these models  quantum integrable. For our supersymmetric
models there is
no obvious way to generalize the zero-curvature condition from the
bosonic models. On the contrary,  
in \cite{holly} it was argued that for bosonic conformal Toda 
theories some conserved currents of higher spin do not have an 
equivalent in the corresponding
supersymmetric theories and it was concluded that these
models are not integrable. 
Whereas this seems to settle the question for conformal Toda theories,
it is an open question, whether or not the supersymmetric
extensions  of affine Toda theories are integrable. 

Affine Toda theories have a unique constant solution as 
a ground state and apart from
$\mathrm{rank}(\g{g})$ fundamental particles, they also admit a set of
soliton solutions  in their spectrum, provided that 
a certain coupling constant $\beta$
is  purely imaginary \cite{sol1, sol3}. There is also a
relationship between
the solitons in affine Toda theories and certain N-body integrable 
systems \cite{nbody1}. 

This paper presents (1,0) supersymmetric extensions of all affine
Toda theories. It is sufficient to consider Toda theories
related to  untwisted, self-dual affine Lie algebras, i.e. those of
the $A-D-E$ series, as all other cases can be related to these.
The  condition for a bosonic massive
sigma model to admit such an extension is that its
scalar potential can be written as the length of a section of a vector
bundle over the sigma model manifold. Employing this result we present
fermionic  extensions of bosonic affine Toda theories with
the following properties:
\begin{description} 
\item[(i)] The classical vacuum is supersymmetric.
\item[(ii)] The mass matrix of the fermions is well defined at
the vacuum.
\item[(iii)] Setting the fermions to zero the usual bosonic affine
Toda theory is recovered.
\item[(iv)] The energy functional is positive semi-definite,
i.e. the inner product of the kinetic terms is positive definite.
\end{description} 
This construction does not render unique supersymmetric
extensions; on the contrary, different ones are explicitly given 
for most bosonic affine Toda theories. 

In section two we give the conditions for bosonic, massive sigma models to
admit (1,0) supersymmetric extensions. In section three we present
(1,0) supersymmetric affine Toda theories  
and in section four we give our
conclusions.

%%%%%%%%%%%%%%%%%%%%%%%%%%%%%%%%%%%%%%%%%%%%%%%%%%%%%%%%%%%%%%%%%%%%%%%%%%

\section{(1,0) supersymmetric massive sigma models}

In this section 
we revise the geometry of a class of (1,0) supersymmetric massive sigma
models with  action
\bqali
&&\hspace{-0.5cm} I=\int \, \mathrm{d^2 x} 
\Big(\partial_{\neq}\phi^i \partial_{=}\phi^j
g_{ij}  + i\lambda_+^i \partial_= \lambda_+^j g_{ij} \label{action}\\ 
&& - \psi^a_-
\partial_= \psi^b_- \delta_{ab}+ m\partial_is_a \lambda^i_+ \psi_-^a -
V(\phi)\Big) \nonumber \ , 
\eqali

\no where $\phi$ is a map from the two-dimensional Minkowski spacetime
$\Sigma$
with  light-cone coordinates $\{x^{\neq}, x^=\}$ into the target
manifold $M$ with metric $g_{ij}$, $1 \leq i \leq \mathrm{dim}(M)$,
$\psi_-^a$ are sections of $\phi^* E
\otimes S_-$ with $E$ a real vector bundle  over $M$, 
$1 \leq a \leq \mathrm{dim}(E)$, and $S_-$ is the bundle
of right handed spinors over $\Sigma$. Furthermore,  $\lambda_+$ are
real chiral fermions, $m$ is a mass parameter, $V$ is the scalar
potential 
\bq
V= \frac{m^2}{4} s_a(\phi) s^a(\phi) \ 
\label{pot}
\eq

\no and $s^a$ are  sections of the vector bundle $E$. 
In (\ref{action})
we  assumed that the sigma model has no torsion, 
that $M$ is flat and that $E$
is trivial, and have chosen the fibre metric of $E$ to be $\delta_{ab}$.
This is the case of interest to us. The action
(\ref{action}) can be
deduced from an off-shell (1,0) superspace formulation for massive
sigma models \cite{mass1} and it is invariant under the (1,0) supersymmetry
transformations
\bq
\delta_{\epsilon} \phi^i = - \frac{i}{2} \epsilon_- \lambda_+^i \, , \ 
\delta_{\epsilon} \lambda_+^i = \frac{1}{2} \epsilon_- \partial_{\neq}
\phi^i \, , \ \delta_{\epsilon} \psi^a_- = \frac{i}{4} 
\epsilon_- m s^a \ , 
\label{susy}
\eq

\no where $\epsilon_-$ is the parameter of the transformations.
Hence we conclude that bosonic 
 massive sigma models admit a (1,0) supersymmetric extension if and
only if their potential can be written as the length of the section of a
vector bundle $E$ over $M$. The zeros of $V$ are the supersymmetric
vacua of the theory. Whereas in the case of (1,0)
supersymmetric models the vector bundle 
$E$ is fixed by the choice of $s$ but otherwise
arbitrary, in the case of a (1,1) supersymmetric extensions $E$ has to be
isomorphic to the tangent bundle of $M$. From 
(\ref{action}) the fermion mass matrix can be read of as 

\bq
\g{M}_{ia}=m\partial_i s_a \ .
\label{mass}
\eq

%%%%%%%%%%%%%%%%%%%%%%%%%%%%%%%%%%%%%%%%%%%%%%%%%%%%%%%%%%%%%%%%%%%%%%%%%%

\section{Supersymmetric affine Toda theories}

In this section we construct  (1,0) supersymmetric extensions of all
Toda theories related to affine algebras. Affine algebras can
be pictorially represented by their generalized Dynkin diagrams. 
(For a  list of the generalized Dynkin diagrams  of all affine
algebras see for 
example \cite[p. 44]{kac}). The 
numbering of the nodes of the generalized Dynkin 
diagrams which we refer to is that of  
E. B. Dynkin \cite{dyn}. For untwisted affine algebras, $\g{g^{(1)}}$, 
the zeroth node represents the highest root $\psi$ of the associated 
simple Lie algebra $\g{g}$ 
and on removal of this node the Dynkin diagram
of $\g{g}$ is recovered with the correct Coexter labels attached to
it. For twisted algebras, $\g{g^{(2)}}$, 
on removal of
the zeroth node the Dynkin diagram of $\g{g}$ is still recovered but
with the wrong Coexter labels attached to it.
Self-dual untwisted affine algebras are associated
to self-dual Lie algebras, which are  $A_r$,
$D_r$, $E_6$, $E_7$ and $E_8$. For all 
other untwisted affine algebras there exist twisted duals, i.e. affine
algebras which are associated to the same  Lie
algebra $B_r$, $C_r$, $F_4$ and $G_2$.

Let us begin with  the action of the bosonic affine Toda theories
associated to the affine algebra $\g{g^{(\epsilon)}}$, $\epsilon=\{1,2\}$, 
\bq
I=\int \, \mathrm{d^2 x} \left( <\partial_{\neq}\phi,
\partial_{=}\phi> - V \right)\ ,
\label{tact}
\eq

\no where $\phi$ is a map from $\Sigma$
into the Cartan subalgebra $\g{h}$ of a simple Lie algebra $\g{g}$,
$<\cdot,\cdot>$ is a metric on $\g{h}$ induced from the invariant
metric on $\g{g^{(\epsilon)}}$ and the potential $V$ is
\bq
V=\frac{m^2}{\beta} \left( \sum_{I=0}^r n_I e^{\beta
\phi_I}-h\right) \ .
\label{v}
\eq

\no The parameters $m$ and $\beta$ are the coupling constants of the
theory,
$\phi_I=\{\phi_0,\phi_i;i=1,2,\ldots,r=\mathrm{rank}(\g{g})\}$, 
$\phi_0=\alpha_0 \cdot \phi$ and 
$\phi_i=\alpha_i\cdot \phi$, where
$\{\alpha_i;i=1,2,\ldots,r\}$ are the simple roots and
$\alpha_0$ is the zeroth root  of the affine algebra
$\g{g^{(\epsilon)}}$. For untwisted affine algebras, $\g{g^{(1)}}$,
$\alpha_0=-\psi$.
The integers $n_I$ are the Coexter labels of $\g{g^{(\epsilon)}}$
defined through $\sum_{I=1}^r n_I\alpha_I=0$ and $h=\sum_{I=1}^r n_I$
is the Coexter number of $\g{g^{(\epsilon)}}$. We observe that $n_0=1$
for all affine algebras.
The  constant term  in the scalar potential  does not affect
the classical theory and hence it is usually neglected. Nevertheless,
we include  $h$ in $V$  in order to render  the vacuum
supersymmetric. 

In order to supersymmetrize  the bosonic action
(\ref{tact}) we have to write the scalar potential 
(\ref{v}) as the sum of squares such
that (i) the vacuum of $V$ at $\phi_i=0$ is supersymmetric and (ii)
the fermions have a well defined mass at the vacuum. We
observe that the scalar potential  only depends on the Coexter labels $n_I$
and not on the generalized Dynkin diagrams themselves. Hence for
this purpose we can treat affine algebras with the same set of Coexter
labels and different Dynkin diagrams, e.g. $B_r^{(1)}$ and $C_r^{(1)}$
for $r>2$, essentially as the same, even though  
they give rise to distinct affine Toda theories. Furthermore, 
in some cases it
is possible to set a subset of the fields $\phi_i$ 
associated to some simple roots of
$\g{g}_1$ in a potential $V_1$ to zero to derive an expression 
for a potential $V_2$ in terms of fields $\tilde{\phi}_i$ of some algebra
$\g{g}_2$.  If we take for example 
$\g{g}_1=F_4$ and $\g{g}_2=G_2$ then the reduced potential of 
$F_4^{(1)}$ can be identified with the one of $G_2^{(1)}$  
by setting  $\phi_1=\tilde{\phi}_1$,
$\phi_2=\tilde{\phi}_2$ and $\phi_3=\phi_4=0$. This identification is
not unique as we can also set $\phi_2=\tilde{\phi}_2$,
$\phi_4=\tilde{\phi}_1$ and $\phi_1=\phi_3=0$. We note that 
the truncation of a set
of Coexter labels does not extend to a truncation of the full affine
Toda theory. On inspection it turns
out that  the set of Coexter labels of any affine
algebra can be derived 
from those of the untwisted, self-dual ones. Thus in the following
we restrict our attention to these affine algebras. 

Let us begin  with  the potential of $D_r^{(1)}$, $r>3$,  proving
inductively that it can be written as the sum of squares such that the
properties
(i) and (ii) are satisfied. For simplicity, we set in the following 
the coupling constants $m$ and $\beta$ to one. Then the  scalar potential of
$D_r^{(1)}$ is 
\bqali
&& \!\!\!\!\!\!\!\!\!\!\!V_{D_r^{(1)}}(\p1, \p2, \ldots,\p{r})
 = e^{\p1} + e^{\p{r-1}} + e^{\p{r}} \\
&& \ + 2 \sum_{i=2}^{r-2} e^{\p{i}} +
e^{-(\p1+\p{r-1}+\p{r})} \prod_{i=2}^{r-2} e^{-2\p{i}} - 2(r-1) \ , 
\nonumber 
\eqali

\no which can be rewritten as 
\bqali
&& \!\!\!\!\!\!\!\!\!\!\!V_{D_r^{(1)}}(\p1, \p2, \ldots,\p{r})
 = \left(e^{-\frac{\p1+\p{r-1}+\p{r}}{2}}\prod_{i=2}^{r-2}e^{\p{i}}-
e^{\frac{\p1}{2}}\right)^2 
+ \left(e^{\frac{\p{r-1}}{2}}-1\right)^2 \label{dr}  \\ && \ +
 \left(e^{\frac{\p{r}}{2}}-1\right)^2 +
2\sum_{i=2}^{r-2} \left(e^{\frac{\p{i}}{2}}-1\right)^2
+2 V_{D_{r-1}^{(1)}}\left(\p2,
\frac{\p3}{2},\ldots,\frac{\p{r}}{2}\right)
\nonumber \ . 
\eqali

\no To complete the inductive proof we express the scalar potential of
$D_4^{(1)}$ as
\bqali
&& \!\!\!\!\!\!\!\!\!\!\!V_{D_4^{(1)}}(\p1, \p2, \p3, \p4)\label{d4}
= e^{\p1} + 2 e^{\p2} +  e^{\p3} +  e^{\p4} +
 e^{-(\p1+ 2\p2 + \p3 + \p4)} - 6   \\ && =
\left(e^{-\frac{\p1 + 2 \p2 + \p3 + \p4 }{2}}-e^{\frac{\p1}{2}}\right)^2 +
 \left(e^{\frac{\p3}{2}}-1\right)^2 +
\left(e^{\frac{\p4}{2}}-1\right)^2  \nonumber \\ && \ 
+2\left(e^{\frac{  \p2}{2}}-1\right)^2
+2\left(e^{-\frac{ 2 \p2 + \p3  + \p4 }{4}}-e^{\frac{\p3}{4}}\right)^2
+4\left(e^{-\frac{ 2 \p2 + \p4 }{8}}-e^{\frac{\p2}{4}}\right)^2 
\nonumber \\ && \ 
+2\left(e^{\frac{ \p4 }{4}}-1\right)^2
+4\left(e^{\frac{ \p4 }{8}}-1\right)^2 
+8\left(e^{\frac{ \p4 }{16}}-e^{-\frac{\p4}{16}}\right)^2 \ . 
\nonumber 
\eqali

The potentials of all twisted and untwisted affine algebras of 
the $A$, $B$, $C$ and $D$ series 
except $A_r^{(1)}$  can be derived from the expressions
(\ref{dr}) and (\ref{d4}) for the potential of $D_r^{(1)}$. 
The case of  $A_r^{(1)}$ was already discussed
in \cite{toda}. The potentials of all remaining affine algebras can be
derived from the ones of  $E_6^{(1)}$,  $E_7^{(1)}$ and  $E_8^{(1)}$. Let us
begin  with the case of  $E_6^{(1)}$:
\bqali
&& \!\!\!\!\!\!\!\!\!\!\! V_{E_6^{(1)}}(\p1, \p2, \p3, \p4, \p5 , \p6) 
= e^{\p1} + 2 e^{\p2}  + 3 e^{\p3}+ 2 e^{\p4} + e^{\p5} + 2 e^{\p6}
\\ && \ + 
e^{-(\p1 + 2 \p2 + 3 \p3 + 2 \p4 + \p5  + 2 \p6 )} -12 \nonumber \\
&& = \left(e^{-\frac{\p1 + 2  \p2+ 3 \p3  + 2 \p4 + \p5 + 2 \p6}{2}}-
e^{\frac{\p3}{2}}\right)^2 
\nonumber \\  && \ +
\sum_{j=1,5}\left(e^{-\frac{\p1+ 2  \p2 + 2 \p3 + 2 \p4  + \p5 + 2 \p6}{4}}-
e^{\frac{\p{j}}{2}}\right)^2 +
8 \left(e^{\frac{\p4 + \p6}{8}}-e^{-\frac{\p4 + \p6}{8}}\right)^2 
\nonumber \\  && \ + 
2e^{-\frac{\p2 + \p3 + \p4 + \p6}{2}}
\left(e^{\frac{\p1 - \p5}{8}}-e^{-\frac{\p1 - \p5}{8}}\right)^2  +
4 \left(e^{-\frac{\p2  + \p3 + \p4 + \p6}{4}}-e^{\frac{\p2 +
\p3}{4}}\right)^2 \nonumber \\ && \ +
 2 \left(e^{\frac{\p3}{2}}-e^{\frac{\p2}{2}}\right)^2 +
2\! \sum_{j=4,6}\left(e^{\frac{\p{j}}{2}}-1\right)^2  +
4\left(e^{\frac{\p4}{4}}-e^{\frac{\p6}{4}}\right)^2 \ . 
\nonumber 
\eqali

\no Similarly, for $E_7^{(1)}$ we find
\bqali
&&\!\!\!\!\!\!\!\!\!\!\! V_{E_7^{(1)}}(\p1, \p2, \p3, \p4, \p5 , \p6, \p7) 
 =2 e^{\p1}  + 3 e^{\p2} + 4 e^{\p3} + 
3 e^{\p4}+ 2 e^{\p5} \\ && 
\ + e^{\p6} +  2 e^{\p7} +
e^{-(2 \p1 + 3 \p2  + 4 \p3+ 3 \p4+ 2 \p5 +\p6 +  2 \p7 )} -18 \nonumber \\ &&
= \left(e^{-\frac{2 \p1 + 3 \p2  + 4 \p3+ 3 \p4+ 2 \p5 +\p6 +  2
\p7}{2}}- e^{\frac{\p7}{2}}\right)^2 \nonumber  \\ && \ +
\sum_{j=2,4}\left(e^{-\frac{2 \p1 + 3 \p2  + 4 \p3+ 3 \p4+ 2 \p5 +\p6 +  
\p7}{4}}-e^{\frac{\p{j}}{2}}\right)^2 \nonumber  \\ && \ +
2 e^{-\frac{ 2 \p1 + 2 \p2  +4 \p3 + 2 \p4 + 2 \p5 + \p6 +\p7}{4}}
\left(e^{\frac{\p2 - \p4}{8}}-e^{\frac{\p4 - \p2}{8}}\right)^2 
 \nonumber  \\ && \ +
2\sum_{j=2,4} \left(e^{-\frac{ 2 \p1+ 2\p2 + 4 \p3 + 2 \p4 + 2 \p5
+\p6 +  \p7 }{8}} -e^{\frac{\p{j}}{2}}\right)^2 \nonumber  \\ && \ +
4 e^{-\frac{ 2\p1 + 4 \p3 +  2\p5 +\p6 +   \p7 }{8}} 
\left(e^{-\frac{\p2 - \p4}{8}}-e^{\frac{\p4 - \p2 }{8}}\right)^2  +
4 \left(e^{\frac{\p3}{2}}-1\right)^2  
\nonumber \\ && \ +
8\left(e^{-\frac{  2 \p1 + 4 \p3 + 2 \p5 + \p6 + \p7 }{16}} -
e^{\frac{\p3}{4}}\right)^2 +
16\left(e^{-\frac{ 2 \p1 + 2 \p5 + \p6 +
\p7}{32}}-e^{\frac{\p1 + \p5}{16}}\right)^2 
\nonumber  \\ && \ +
2 \left(e^{\frac{\p1}{2}}-e^{\frac{\p5}{2}}\right)^2 +
\sum_{b=4,8} b \left(e^{\frac{\p1+ \p5}{b}}-1\right)^2  +
\left(e^{\frac{\p6}{2}}-e^{\frac{\p7}{2}}\right)^2 \nonumber  \\ && \ +
 \sum_{b=2,4,8,16} b \left(e^{\frac{\p6 + \p7}{b}}-1\right)^2  +
32\left(e^{-\frac{\p6 + \p7}{64}}-e^{\frac{\p6 + \p7}{64}}\right)^2  
\nonumber \ . 
\eqali

\no Finally, for $E_8^{(1)}$ we rewrite the potential as 
\bqali
&& \!\!\!\!\!\!\!\!\!\!\!V_{E_8^{(1)}}(\p1, \p2, \p3 \p4, \p5 , \p6 , \p7,
\p8)  
= 2 e^{\p1} + 3 e^{\p2} + 4 e^{\p3} + 5 e^{\p4}+ 6 e^{\p5} \\ && \  
+ 4 e^{\p6}  + 2 e^{\p7} + 3 e^{\p8}  + 
e^{-(2 \p1+ 3 \p2  + 4 \p3  + 5 \p4 + 6 \p5+ 4 \p6 + 2 \p7 + 3 \p8)} 
-30 \nonumber \\ &&
= \left(e^{-\frac{2 \p1 + 3 \p2+ 4 \p3 + 5 \p4 + 6 \p5+ 4 \p6  + 
2 \p7 + 3 \p8 }{2}}-e^{\frac{\p4}{2}}\right)^2 \nonumber \\ && \ +
\sum_{j=2,8}\left(e^{-\frac{2 \p1 + 3 \p2+ 4 \p3 + 4 \p4 + 6 \p5+ 4 \p6  + 
2 \p7 + 3 \p8 }{4}}-e^{\frac{\p{j}}{2}}\right)^2 \nonumber \\ && \ +
2 e^{-\frac{\p1 + \p2   + 2 \p3 + 2 \p4 + 3 \p5+ 2 \p6 + 
 \p7+  \p8}{2}}\left(e^{\frac{\p2 - \p8}{8}}-
e^{-\frac{\p8 - \p2}{8}}\right)^2 \nonumber \\ && \ +
4\left(e^{-\frac{ \p1 +  \p2+ 2 \p3 + 2 \p4 + 3 \p5+ 2 \p6 +  \p7  
+  \p8 }{4}}-e^{\frac{\p5}{2}}\right)^2 \nonumber \\ && \ +
4\sum_{j=3,6}\left(e^{-\frac{ \p1 +  \p2+ 2 \p3 + 2 \p4 +  \p5+ 2 \p6 +  \p7  
+  \p8 }{8}}-e^{\frac{\p{j}}{2}}\right)^2 \nonumber \\ && \ +
8 e^{-\frac{ \p1 +  \p2  + 2 \p4 + \p5 +  \p7 + \p8 }{8}} 
\left(e^{\frac{\p3-\p6}{8}}-e^{\frac{\p6-\p3}{8}}\right)^2
\nonumber \\ && \ +
16\left(e^{-\frac{\p1 + \p2 + 2 \p4 + \p5+ \p7 + \p8 }{16}}-
e^{\frac{\p4}{8}}\right)^2 +
4 \left(e^{\frac{\p4}{2}}-1\right)^2 \nonumber \\ && \ +
8 \left(e^{\frac{\p4}{4}}-1\right)^2  + 
32\left(e^{-\frac{\p1 + \p2 + \p5 + \p7 + \p8}{32}}-
e^{\frac{\p5}{32}}\right)^2 \nonumber \\ && \ +
\sum_{b=2,4,8,16} b \left(e^{\frac{\p5}{b}}-1\right)^2 +
2 \left(e^{\frac{\p1}{2}}-e^{\frac{\p7}{2}}\right)^2 +
2 \left(e^{\frac{\p2}{2}}-e^{\frac{\p8}{2}}\right)^2 \nonumber \\ && \ +
4 \left(e^{\frac{\p1 + \p7}{4}}-e^{\frac{\p2+ \p8}{4}}\right)^2 +
\sum_{b=8,16,32} b \left(e^{\frac{\p1 + \p2+ \p7  + \p8}{b}}-1\right)^2  
\nonumber \\ && \  +
64  \left(e^{-\frac{\p1 + \p2 + \p7 + \p8}{64}}- 
e^{\frac{\p1 + \p2 + \p7 + \p8}{64}}\right)^2
\nonumber  \ . 
\eqali

Let us consider as an example of our construction the
untwisted affine algebra
$G_2^{(1)}$. Its scalar potential can be derived
from  the ones associated to any of the algebras of the $E$ series. We
choose in the following the  truncation of $E_6^{(1)}$ for which the
dimension of the vector bundle $E$ is minimal and reintroduce the
coupling constants $m$ and $\beta$. In this case the appropriate
 truncation of fields is $\phi_2=\tilde{\phi}_1$,
$\phi_3=\tilde{\phi}_2$ and $\phi_1=\phi_4=\phi_5=\phi_6=0$. 
Then we find  the  section $s^a$ of $E$ for $V_{G_2^{(1)}}$ as
\bqali
&& \hspace{-3mm}\{s^a\}=\frac{1}{\beta}\Big\{e^{-\beta \frac{2\p1 + 3
\p3}{2}}-e^{\beta \frac{\p1}{2}}, \label{g2} \\
&& \sqrt{2}\left(e^{-\beta \frac{\p1 + \p2 }{2}}-1\right), 
2\left(e^{\beta \frac{\p1 + \p2}{4}}- e^{-\beta \frac{\p1 + \p2}{4}}\right), 
\sqrt{2}\left(e^{\beta\frac{\p1}{2}}-e^{\beta\frac{\p2}{2}}\right)\Big\}
\nonumber \ ,
\eqali

\no where $a=\{1,2,3,4\}$. The vector bundle $E$ is four-dimensional
and it is clear that we should add another  four chiral fermions
$\{\psi_-^a;a=1,2,3\}$ to the two fermions of opposite chirality 
 $\{\lambda_+^i;i=1,2\}$. The (1,0) supersymmetric action for the
affine Toda theory based on $G_2^{(1)}$ can be found by substituting (\ref{g2})
in (\ref{action}). In this case the fermion mass matrix is a $2\times3$ matrix,
which takes at the vacuum, $\phi=0$, the following form
\bq
\g{M}|_{\phi=0} = m \left(\begin{array}{cccc}
-\frac{3}{2} &-\frac{3}{\sqrt{2}} & 1 & \frac{1}{\sqrt{2}} \\
-\frac{3}{2} &-\frac{3}{\sqrt{2}} & 1 & - \frac{1}{\sqrt{2}}
\end{array}\right) \ . 
\eq

\no This matrix has rank two and so there are two Majorana fermions
constructed from the eight pairs $\{\lambda_+,\psi_-\}$ of
Majorana-Weyl fermions which have non-zero mass at the supersymmetric
vacuum.

%%%%%%%%%%%%%%%%%%%%%%%%%%%%%%%%%%%%%%%%%%%%%%%%%%%%%%%%%%%%%%%%%%%%%%%%%%

\section{Concluding remarks}

In this paper we have constructed (1,0) supersymmetric extensions of
all bosonic affine Toda theories.  Our results are  based 
on  supersymmetric  massive sigma models, i.e.  the
bosonic Toda theories are extended  without reference to superalgebras. 
The advantage of our method is that these models are unitary and their
supersymmetry  is unbroken. 
Furthermore, the  mass matrix of the fermions is
well defined at the vacuum,
 and  the usual bosonic affine Toda theories is recovered
on setting the fermions to zero.  

It  is  sufficient to consider affine Toda theories 
based on  affine algebras
of the $A-D-E$ series, since all the other cases can be deduced from
these.  We found that the supersymmetrization of affine  Toda
theories  is in general  not unique. On the contrary, there
are as many extensions of the bosonic  theories
as there are ways to write their scalar
potential as the sum of squares. 

In this context it would be
interesting to find a geometric interpretation of the vector bundle
$E$, as the different extensions of affine Toda theories 
correspond to different choice of a
section of the vector bundle
$E$. Another   question arises as to whether or not our
supersymmetric Toda theories share the integrability property with
their bosonic counterparts. The
conventional methods of proving integrability through Lax pairs or zero
curvature conditions  have 
no obvious generalization from the bosonic to the fermionic Toda theories. 
Whereas there are arguments why similar extensions of conformal
Toda theories should not be integrable, this question remains open
for the supersymmetric affine Toda theories constructed in
this paper.

\vskip 1cm
\noindent{\bf Acknowledgments:} The author  would like thanks G. Papadopoulos 
 for helpful discussions and advice and the EPSRC and the German
 National Foundation for financial support.
\vskip 1cm

%%%%%%%%%%%%%%%%%%%%%%%%%%%%%%%%%%%%%%%%%%%%%%%%%%%%%%%%%%%%%%%%%%%%%%%%%%

%%%%%%%%%%%%%%%%%%%%%%%%%%%%%%%%%%%%%%%%%%%%%%%%%%%%%%%%%%%%%%%%%%%%%%%%%%

\end{document}